\documentclass{article}
\usepackage{amsfonts}

\newcommand{\beq}{\begin{equation}}
\newcommand{\eeq}{\end{equation}}
\newcommand{\dpl}{\displaystyle}

\begin{document}

\title{Deformations of the root systems
and 
new solutions to generalised WDVV equations}

\maketitle

\begin{center}

{\bf A.P.Veselov }

\bigskip

{\it Department of Mathematical Sciences, Loughborough University,
Loughborough, Leicestershire, LE 11 3TU, UK
}

\bigskip

{\it Landau Institute for Theoretical Physics, Kosygina 2,

 Moscow, 117940,  Russia

\bigskip

e-mail: A.P.Veselov@lboro.ac.uk,
}

\end{center}

\bigskip

\bigskip

{\small  {\bf Abstract.}
A special class of solutions to the generalised WDVV equations related to 
a finite set of covectors is investigated. Some geometric conditions on such a set
which guarantee that the corresponding function satisfies WDVV equations are found
($\vee$-conditions). These conditions are satisfied for all root systems and their
special deformations discovered in the theory of the Calogero-Moser systems by 
O.Chalykh, M.Feigin and the author. This leads to the new solutions for the generalized 
WDVV equations.}

\bigskip

\section*{Introduction.}

The generalised WDVV (Witten-Dijgraaf-Verlinde-Verlinde) equations are
 the following overdetermined system of nonlinear differential equations:
\beq
F_iF_k^{-1}F_j=F_jF_k^{-1}F_i, \quad i,j,k=1,\ldots,n,
\label{wdvv}
\eeq
where $F_m$ is the $n\times n$ matrix constructed from the third partial
 derivatives of the unknown function $F=F(x^1,\ldots,x^n)$:
\beq
\label{f}
(F_m)_{pq}=\frac{\partial ^3 \, F}{\partial x^m \partial x^p \partial x^q},
\eeq
In this form these equations have been presented by A.Marshakov,A.Mironov and A.Morozov, 
who showed that the Seiberg-Witten prepotential in $N=2$ four-dimensional 
supersymmetric gauge theories satisfies this system \cite{MMM}. 
Originally these equations appeared in a slightly different form as associativity 
conditions in topological field theory (see \cite{D}). 

In the recent paper \cite{MG} it has been shown that 
for any root system ${\cal R} \in {\bf R}^n$ of a semisimple Lie algebra of rank $n$ 
the function
\beq
\label{F}
F(x)=\frac{1}{4}\sum\limits_{\alpha \in {\cal R}} (\alpha,x)^2  {\rm log} \, 
(\alpha,x)^2
\eeq
satisfies the WDVV equations (\ref{wdvv},\ref{f}).

 Our first observation is that this is actually true for all the Coxeter configurations
related to any finite reflection  group.

 The second, main observation is that the function (\ref{F}) satisfies WDVV equation
also for certain deformations of the root systems  discovered by O.Chalykh, M.Feigin 
and the author \cite{ChFV1,ChFV2,ChFV3}. The corresponding families of the new solutions
 to WDVV equations have the form
\beq
\label{NF1}
F=\sum\limits_{i<j}^{n} (x_i-x_j)^2 \, {\rm log} \, (x_i-x_j)^2 +
\frac{1}{m} \sum\limits_{i=1}^nx_i^2 \,{\rm log} \, x_i^2
\eeq
with an arbitrary real value of the parameter $m$ and
\beq
\label{NF2}
\begin{array}{c}
F= k\sum\limits_{i<j}^{n}\left[ (x_i+x_j)^2 \, {\rm log} \, (x_i+x_j)^2 +
(x_i-x_j)^2  {\rm log} \, (x_i-x_j)^2\right]+ \\
+ \sum\limits_{i=1}^n \left[ (x_i+x_{n+1})^2 {\rm log} \, (x_i+x_{n+1})^2+
 (x_i-x_{n+1})^2 {\rm log} \, (x_i-x_{n+1})^2 \right]+ \\ 
+4m \sum\limits_{i=1}^n x_i^2 \,{\rm log} \, x_i^2+
4l x_{n+1}^2 \,{\rm log} \, x_{n+1}^2,
\end{array}
\eeq
where the real parameters $k, m, l$ satisfy the only relation
\beq
k (2l +1)=2m +1.
\label{rel}
\eeq
When $m =1$ the formula (\ref{NF1})  gives the well-known solution to WDVV equations, 
corresponding to the leading perturbative approximation to the exact Seiberg-Witten
prepotential for the gauge group $SU(n+1)$ (see \cite{MMM}).
For the general $m$ it corresponds to the deformation $A_{n}(m)$ of the root
system $A_n$ related to the Lie algebra $su(n+1)$ (see \cite{ChFV1} and below).
 The formula (\ref{NF2}) with $k=m=l=1$ correspond to the root system
$C_{n+1}$, the general case - to its deformation  $C_{n+1}(m,l)$
(see \cite{ChFV3}). 

Actually I will describe some geometric conditions ($\vee$-conditions) on the set
of the covectors $\alpha$, which guarantee that the corresponding function (\ref{F})
satisfies the generalised WDVV equations. The roots systems and their deformations
mentioned above satisfy these conditions. The classification of all $\vee$-systems
seems to be an interesting open problem.

\section{$\vee$-systems and particular solutions to WDVV equations.}

 Let's first remind the following observation from \cite{MMM,MM}  that WDVV
equations (\ref{wdvv}), (\ref{f}) are equivalent to the equations
\beq
F_iG^{-1}F_j=F_jG^{-1}F_i, \quad i,j=1,\ldots ,n,
\label{fgf}
\eeq
where $G=\sum\limits_{k=1}^n\eta^kF_k$ is any particular invertible linear combination of $F_i$
with the coefficients, which may depend on $x$. Introducing the matrices $\check F_i=
G^{-1}F_i$ one can rewrite (\ref{fgf}) as the commutativity relations 
\beq
\left[ \check F_i, \check F_j \right] =0, \quad i,j=1,\ldots ,n,
\label{com}
\eeq
Let us consider now the following particular class of the solutions to these equations.

Let $V$ be a real linear vector space of dimension $n$, $V^*$ be its dual space consisting of 
the linear functions on $V$ (covectors),  $\mathfrak{A}$ be a finite
 set of noncollinear covectors $\alpha \in V^*$.

Consider the following function on $V$:
\beq
F^{\mathfrak{A}}=\sum\limits_{\alpha \in \mathfrak{A}} (\alpha,x)^2 \, {\rm log} \, 
(\alpha,x)^2,
\label{mF}
\eeq
where $(\alpha,x)=\alpha(x)$ is the value of covector $\alpha \in V^*$ on a vector $x\in V$. 
For any basis $e_1,\ldots, e_n$ we have the corresponding coordinates
 $x^1,\ldots , x^n$
in $V$ and the matrices $F_i$ defined according to  (\ref{f}). In a more invariant form
 for any vector $a \in V$ one can define the matrix 
\beq
\label{fa}
F_a=\sum\limits_{i=1}^n a^iF_i.
\eeq
By a straightforward calculation one can check the following

\smallskip

{\bf Lemma.} {\it  For a function (\ref{mF})  $F_a$  is the matrix 
of the following bilinear form on  $V$
\beq
\label{mFa}
F_a^{\mathfrak{A}}=\sum\limits_{\alpha \in \mathfrak{A}} \frac{(\alpha,a)}{(\alpha,x)}
\alpha\otimes \alpha,
\eeq
where $\alpha\otimes \beta (u,v)=\alpha (u)\beta (v)$ for any $u,v \in V$ and $\alpha, \beta \in V^*$.}

Now let us choose $G^{\mathfrak{A}}$ as $F_x^{\mathfrak{A}}$:
$$
G^{\mathfrak{A}}=\sum\limits_{i=1}^n x^iF_i.
$$
Then, according to the lemma $G^{\mathfrak{A}}$ is a matrix of the bilinear form
\beq
G^{\mathfrak{A}}=\sum\limits_{\alpha \in \mathfrak{A}} \alpha\otimes \alpha,
G^{\mathfrak{A}} (u,v)=\sum\limits_{\alpha \in \mathfrak{A}} (\alpha,u)(\alpha,v).
\label{mG}
\eeq
Notice that $G^{\mathfrak{A}}$ does not depend on $x$.

We will assume that $G^{\mathfrak{A}}$ is non-degenerate, which in the real case means 
that the covectors
 $\alpha \in \mathfrak{A}$ generate $V^*$. This means  that the natural linear mapping
$\varphi_\mathfrak{A} : V\rightarrow V^*$  defined by the formula 
$$
(\varphi_\mathfrak{A}(u),v)=G^{\mathfrak{A}} (u,v), \, u,v \in V
$$
is invertible. We will denote  $\varphi_\mathfrak{A}^{-1}(\alpha),\,  \alpha \in V^*$ as
$\alpha^{\vee}$. By definition, for any $v \in V$
\beq
(\alpha,v)=\sum\limits_{\alpha \in \mathfrak{A}}(\alpha, \alpha^{\vee})(\alpha,v).
\label{vee}
\eeq
Now let us define the operator
\beq
\check F_a^{\mathfrak{A}}=\sum\limits_{\alpha \in \mathfrak{A}} \frac{(\alpha,a)}
{(\alpha,x)}\alpha^\vee \otimes \alpha
\label{oper}
\eeq
According to (\ref{com})
 the WDVV equations (\ref{wdvv},\ref{f}) for the function (\ref{mF}) are equivalent to
\beq
\label{ab}
\left[\check F_a^\mathfrak{A}, \check F_b^\mathfrak{A} \right] =0
\eeq
for any $a,b \in V$. A simple calculation  shows that (\ref{ab}) can be rewritten as 
\beq
\label{sc}
\sum\limits_{\alpha \ne \beta, \alpha,\beta \in \mathfrak{A}}
\frac{G^\mathfrak{A} (\alpha^\vee, \beta^\vee)B_{\alpha,\beta}(a,b)}
{(\alpha,x)(\beta,x)}\alpha\wedge \beta \equiv 0,
\eeq
where 
$$
\alpha\wedge \beta=\alpha\otimes \beta-\beta \otimes \alpha
$$ 
and
$$
B_{\alpha,\beta}(a,b)=\alpha\wedge \beta(a,b)=\alpha(a)\beta(b)-\alpha(b)\beta(a).
$$
Thus the WDVV equations for the function (\ref{mF}) are equivalent to the conditions
(\ref{sc}) to be satisfied for any $x,a,b \in V$.

 Notice that WDVV equations (\ref{wdvv},\ref{f}) and, therefore, the conditions (\ref{sc}) 
are obviously satisfied
for any two-dimensional configuration $\mathfrak{A}$. This fact and the structure 
of the relation (\ref{sc}) motivate the following notion of the $\vee$-systems.

  Remind first that for a pair of  bilinear forms $F$ and $G$ on the vector space $V$
 one can define an eigenvector $e$ as the kernel of the bilinear form $F-\lambda G$
 for a proper $\lambda$:
$$
(F-\lambda G) (v,x)=0
$$
 for any $v \in V$. When $G$ is non-degenerate $e$ is the eigenvector of the corresponding operator 
$\check F=G^{-1} F$:
$$
\check F(e)=G^{-1}F(e)=\lambda e.
$$
Now let $\mathfrak{A}$ be as above any finite set of non-collinear covectors
$\alpha \in V^*$, $G=G^\mathfrak{A}$ be the corresponding bilinear form (\ref{mG}),
which is assumed to be non-degenerate, $\alpha^\vee$ are defined by
(\ref{vee}). Define now for any two-dimensional plane $\Pi \subset V^*$ a form
\beq
G^\mathfrak{A}_\Pi (x,y)= \sum\limits_{\alpha \in \Pi \cup \mathfrak{A}} (\alpha,x)(\alpha,y).
\label{GP}
\eeq

\smallskip

{\bf Definition.} {\it We will say that $\mathfrak{A}$ satisfies the $\vee$-conditions 
if for any plane $\Pi \in V^*$ the vectors $\alpha^\vee, \, \alpha \in \Pi \cup \mathfrak{A}$
are the eigenvectors of the pair of the forms $G^\mathfrak{A}$ and $G^\mathfrak{A}_\Pi$. 
In this case we will call  $\mathfrak{A}$ as $\vee$-system.}

\smallskip

The $\vee$-conditions can be written explicitly as
\beq
\sum\limits_{\beta \in \Pi \cap \mathfrak{A}}
\beta(\alpha^\vee)\beta^\vee=\lambda \alpha^{\vee},
\label{expl}
\eeq
for any $\alpha \in \Pi \cap \mathfrak{A}$ and some $\lambda$, which may depend 
on $\Pi$ and $\alpha$.

If the plane $\Pi$ contains 
no more that one vector from $\mathfrak{A}$ then this condition is obviously satisfied,
so the $\vee$-conditions should be checked only for a finite number of planes $\Pi$. 

If the plane $\Pi$
contains only two covectors $\alpha$ and $\beta$ from $\mathfrak{A}$ then the 
condition (\ref{expl}) means that $\alpha^\vee$ and $\beta^\vee$ are orthogonal with respect 
to the form $G^\mathfrak{A}$:
$$
\beta(\alpha^\vee) = G^\mathfrak{A}( \alpha^\vee,\beta^\vee)=0.
$$

If the plane $\Pi$ contains more that two covectors from $\mathfrak{A}$ this condition
means that $G$ and $G_\Pi$ restricted to the plane $\Pi^\vee \subset V$ are
proportional:
\beq
\label{restr}
\left. G_\Pi \right|_{\Pi^\vee}=\lambda (\Pi) \left. G \right|_{\Pi^\vee}
\eeq


\smallskip

{\bf Theorem 1.} {\it Let $\mathfrak{A}$ be any $\vee$-system, then the function (\ref{mF})
satisfies the WDVV equations (\ref{wdvv}).}

\smallskip

{\it Proof.}  It is enough to prove that 
$$
\sum\limits_{\alpha,\beta \in \Pi \cap \mathfrak{A}}
\frac{G^\mathfrak{A} (\alpha^\vee, \beta^\vee) B_{\alpha,\beta}(a,b)}
{(\alpha,x)(\beta,x)}\alpha\wedge \beta \equiv 0
$$
 for any plane $\Pi \in V^*$. When $ \Pi \cap \mathfrak{A}$ consists only
of two covectors $\alpha$ and $\beta$ this follows from
$G^\mathfrak{A} (\alpha^\vee, \beta^\vee)=0$, which is the $\vee$-condition in this case.
If  $\Pi \cap \mathfrak{A}$ consists of more that two covectors this relation 
is proportional to the corresponding relation for the function
$$
\left. F^\mathfrak{A}_\Pi\right|_{\Pi^\vee}=
\sum\limits_{\alpha \in \mathfrak{A}\cap \Pi} (\alpha,x)^2  \left. {\rm log} \, 
(\alpha,x)^2 \right|_{\Pi^\vee}.
$$
Since $\Pi^\vee$ is two-dimensional this is obviously satisfied.

\section{Examples of $\vee$-systems: root systems and their deformations.}

Let $V$ be now Euclidean vector  space with a scalar product $(\, ,\,) $, 
and $G$ be any irreducible finite group generated by orthogonal reflections 
with respect to some hyperplanes (Coxeter groups \cite{Burb}).  Let ${\cal R}$ be a set of normal
vectors to the reflection hyperplanes of $G$. We will not fix the length of the normals but
assume that ${\cal R}$ is invariant under the natural action of $G$ and contains exactly two normal 
vectors for any such hyperplane. Let us choose from each such pair of
vectors one of them and form the system ${\cal R}_{+}$:
$$
{\cal R}={\cal R}_{+}\cup ( -{\cal R}_{+}).
$$
Usually  ${\cal R}_{+}$ is chosen simply by taking from ${\cal R}$ vectors which are positive 
with respect to some linear form on $V$. We will call a system ${\cal R}_{+}$ as 
{\it Coxeter system} and  the vectors from ${\cal R}_{+}$ as {\it roots}.

\smallskip

{\bf Theorem 2.} {\it  Any Coxeter  system ${\cal R}_{+}$ is a $\vee$-system.}

\smallskip

 Proof is very simple. First of all the form (\ref{mG})   in this case is 
proportional  to the euclidean structure on $V$ because it is invariant under $G$ 
and $G$ is irreducible. By the same reason this is true for the form $G_\Pi$ (\ref{GP})
if the plane $\Pi$ contains more than two roots from ${\cal R}_{+}$. When $\Pi$
contains only two roots they must be orthogonal and therefore satisfy
 $\vee$-conditions.

 {\bf Corollary.} {\it For any Coxeter system ${\cal R}_{+}$ the function
\beq
\label{Cor}
F=\sum\limits_{\alpha \in {\cal R}_{+}} \,  (\alpha,x)^2  \, {\rm log} \, 
(\alpha,x)^2
\eeq
satisfy WDVV equations (\ref{wdvv}), (\ref{f}).}

{\it Remark}. The root systems of any semisimple Lie algebra are the particular
 examples of the Coxeter systems. In this case this result has been proven
 in \cite{MG}. Notice that even when $G$ is a Weyl group of some Lie algebra
our formula (\ref{Cor}) in general gives more solutions since we have not fixed 
the length of the roots.

\smallskip

 Remarkably enough that the theorem 2 is true also for the following 
deformations of the root systems found in the theory of the generalised Calogero-Moser 
systems in \cite{ChFV1,ChFV2,ChFV3}.

 Let us make first the following remark. One can consider the class of functions 
related to a formally more general situation when the
 covectors $\alpha$ have also some prescribed multiplicities $\mu_\alpha$
\beq
F^{(\mathfrak{A},\mu)}=\sum\limits_{\alpha \in \mathfrak{A}} \mu_\alpha (\alpha,x)^2 
\, {\rm log} \, (\alpha,x)^2.
\label{frakF}
\eeq
But it is easy to see that this actually will give no new solutions because 
$F^{(\mathfrak{A},\mu)}=F^{\tilde \mathfrak{A}}+$ quadratic terms, where  
$\tilde \mathfrak{A}$ consists of covectors $\sqrt{\mu_\alpha} \alpha$.

  The following configurations $A_n(m)$ and $C_{n+1}(m,l)$ have been introduced
in \cite{ChFV1,ChFV2,ChFV3}. They consist of the following vectors in ${\bf R}^{n+1}$:
$$
A_n(m)=
\left\{
\begin{array}{lll}
e_i - e_j, &  1\le i<j\le n, & {\rm with \,\, multiplicity \,\,}   m,\\
e_i - \sqrt{m}e_{n+1}, &  i=1,\ldots ,n  & {\rm with \,\, multiplicity \,\,}  1,
\end{array}
\right.
$$
and
$$
C_{n+1}(m,l) =
\left\{
\begin{array}{lll}
e_i\pm e_j, &  1\le i<j\le n, &  {\rm with \,\, multiplicity \,\,}   k,\\
2e_i, &  i=1,\ldots ,n  & {\rm with \,\, multiplicity \,\,}   m,\\
e_i\pm \sqrt{k}e_{n+1}, & i=1,\ldots ,n  &  {\rm with \,\, multiplicity \,\,}  1,\\
2\sqrt{k}e_{n+1} & {\rm with \,\, multiplicity \,\,}  l,\\
\end{array}
\right.
$$
where $k = \frac{2m+1}{2l+1}$.

When all the multiplicities are integer the  corresponding  generalisation of
Calogero-Moser system is algebraically integrable, but usual integrability 
holds for any value of multiplicities (see \cite{ChFV1,ChFV2,ChFV3}). 

Notice that when $m=1$ the first configuration coincides with the classical root system
of type $A_n$ and when $k=m=l=1$ the second configuration is the root system of type $C_{n+1}$.
So these families can be considered as the special deformations of these roots systems.

\smallskip

{\bf Theorem 3.} {\it For the deformed root systems $A_n(m)$ and $C_{n+1}(m,l)$  with
arbitrary $m$ and $l$ the function (\ref{frakF}) satisfies WDVV equation.
}

\smallskip

This follows from the simple check that the sets
$$
\tilde A_n(m)=
\left\{
\begin{array}{ll}
\sqrt{m} \, (e_i - e_j ), &  1\le i<j\le n,\\
\\
e_i - \sqrt{m}e_{n+1}, & i=1,\ldots ,n
\end{array}
\right.
$$
and
$$
\tilde C_{n+1}(m,l) =
\left\{
\begin{array}{ll}
\sqrt{k} e_i\pm \sqrt{k} e_j, &    1\le i<j\le n\\
\\
2\sqrt{m} e_i, & i=1,\ldots ,n\\
\\
e_i\pm \sqrt{k}e_{n+1}, & i=1,\ldots ,n\\
\\
2\sqrt{kl}e_{n+1}, & \\
\end{array}
\right.
$$
with $k=\frac{2m+1}{2l+1}$ satisfy the $\vee$-conditions. Making a suitable linear transformation 
one can rewrite these families in the following, more convenient way:
$$
\tilde A_n(m)=
\left\{
\begin{array}{ll}
e_i - e_j , &  1\le i<j\le n,\\
\\
\frac{\dpl 1}{\dpl \sqrt{m}}e_{i}, & i=1,\ldots ,n
\end{array}
\right.
$$
and
$$
\tilde C_{n+1}(m,l) =
\left\{
\begin{array}{ll}
\sqrt{k} (e_i\pm  e_j), & 1\le i<j\le n, \\
\\
2\sqrt{m} e_i, &i=1,\ldots ,n\\
\\
e_i\pm e_{n+1}, & i=1,\ldots ,n\\
\\
2\sqrt{l}e_{n+1}, & \\
\end{array}
\right.
$$
where again $k=\frac{2m+1}{2l+1}$. 

Corresponding functions $F$ have the form (\ref{NF1}), (\ref{NF2}) written in the Introduction.
\smallskip

 {\bf Corollary}. {\it  The functions $F$ given by the formulas (\ref{NF1}), (\ref{NF2})
 satisfy WDVV equations.}

\smallskip

At the moment I have no satisfactory explanation why the deformed root systems arisen in the
theory of the generalised Calogero-Moser problems turned out to be $\vee$-systems.
It may be that it is a common geometrical property of all the so-called locus configurations
\cite{ChFV3}. In this connection I'd like to mention that $\vee$-systems can be naturally defined 
in a complex vector space. All this certainly deserves further investigation.


\begin{thebibliography}{99}
\bibitem{MMM}
A.Marshakov, A.Mironov, and A.Morozov
{\it WDVV-like equations in $N=2$ SUSY Yang-Mills theory.}
Phys.Lett. B, {\bf 389} (1996), 43-52, hep-th/9607109.
\bibitem{D}
B.Dubrovin
{\it Geometry of 2D topological field theories.}
Nucl.Phys., {\bf B352} (1992), 627, hep-th/9407018.
\bibitem{MG}
R.Martini, P.K.H.Gragert
{\it Solutions of WDVV equations in Seiberg-Witten theory from root systems.}
J.of Nonlinear Math.Physics, {\bf 6} (1) (1999), 1-4.
\bibitem{ChFV1}
A.P. Veselov, M.V. Feigin, O.A. Chalykh
{\it New integrable deformations of quantum Calogero - Moser problem.}
Usp. Mat. Nauk {\bf 51} (3) (1996), 185--186.
\bibitem{ChFV2}
O.A. Chalykh, M.V. Feigin, A.P. Veselov
{\it New integrable generalizations of Calogero-Moser quantum problem.}
J. Math. Phys {\bf 39} (2) (1998), 695--703.
\bibitem{ChFV3}
O.A. Chalykh, M.V. Feigin, A.P. Veselov.  {\it Multidimensional Baker-Akhiezer
Functions and Huygens' Principle.} Submitted to Commun. Math. Physics, hep-th/9902141.
\bibitem{MM}
A.Marshakov, A.Mironov, A.Morozov
{\it WDVV equations from algebra of forms.}
Mod.Phys.Lett {\bf A12} (1997) 773, hep-th/9701014.
\bibitem{Burb}
N. Bourbaki, {\it Groupes et alg\`ebres de Lie.} Chap.~VI, Masson, 1981

\end{thebibliography}
\end{document}